\documentclass[12pt]{article}
\pdfoutput=1

\usepackage{epsfig,a4wide,amsmath,amssymb,amsfonts,mathrsfs,amsthm}
\usepackage{mathtools}
\usepackage{xcolor,yfonts}
\usepackage{graphicx}
\usepackage{subcaption}
\usepackage[utf8]{inputenc}
\usepackage{siunitx}
\usepackage{rotating}
\usepackage{hyperref}
\usepackage[normalem]{ulem}
\usepackage{cite}
\usepackage{float}

\newcommand{\be}{\begin{equation}}
\newcommand{\ee}{\end{equation}}
\numberwithin{equation}{section}

\newcommand{\ignore}[1]{}

\def\a{\alpha}
\def\b{\beta}

\begin{document}

\begin{titlepage}

\begin{center}

~\\[2cm]

{\LARGE A new energy inequality in AdS}

~\\[0.5cm]

{\fontsize{14pt}{0pt} Gary~T.~Horowitz${}^{\diamond}$,  Diandian~Wang${}^{\dagger}$, Xiaohua~Ye${}^{\sharp}$}

~\\[0.1cm]

\it{ ${}^{\diamond}$  Department of Physics, University of California, Santa Barbara, CA 93106}

~\\[0.05cm]

\it{ ${}^{\dagger}$ Center for the Fundamental Laws of Nature,}
\\
\it{Harvard University, Cambridge, MA 02138}

~\\[0.05cm]

\it{ ${}^{\sharp}$ Department of Applied Mathematics and Theoretical Physics,}
\\
\it{University of Cambridge,
Cambridge CB3 0WA, UK}

\end{center}

  \vspace{60pt}

\noindent

We study time symmetric initial data for asymptotically AdS spacetimes with conformal boundary containing a spatial circle. Such $d$-dimensional initial data sets can contain $(d-2)$-dimensional minimal surfaces if the circle is contractible. We compute the minimum energy of a large class of such initial data as a function of the area $A$ of this minimal surface. The statement $E \ge E_{min}(A)$ is analogous to the Penrose inequality which bounds the energy from below by a function of the area of a $(d-1)$-dimensional minimal surface.

\vfill

    \noindent

  \end{titlepage}

   \newpage

\tableofcontents
\baselineskip=16pt

\section{Introduction}

One of the major milestones in the development of general relativity was the proof of the positive energy theorem. This was first established for asymptotically flat spacetimes \cite{Schoen:1979zz,Witten:1981mf}, and later generalized to asymptotically anti-de Sitter (AdS) spacetimes with the standard $S^{d-1}\times R$ boundary conditions \cite{Gibbons:1983aq,Wang:2001,Maerten:2006}. However it was noticed that if the conformal AdS boundary contains a spatial circle, there are solutions with energy lower than AdS. The (negative energy) AdS soliton was conjectured to be the lowest energy solution with $S^1 \times R^{d-1}$ boundary \cite{Horowitz:1998ha}, and this has recently been proven \cite{Brendle:2024} (for time symmetric,  topologically $R^d$, initial data).

An interesting consequence of {(spatially compact)} conformal boundaries with a spatial circle is that the initial data always contains a co-dimension two minimal surface when this circle is contractible. We investigate the minimum energy of initial data as a function of the area of this minimal surface. We focus on the vacuum Einstein equation with a negative cosmological constant $\Lambda < 0$. {As one expects that minimal energy solutions are static, or at least have a time symmetric surface, we will focus on initial data with time {reflection} symmetry.} The only constraint on $d$-dimensional, time symmetric initial data is $\mathcal{R} = 2\Lambda$, where $\mathcal{R} $ is the scalar curvature of the spatial metric. {One also expects them to have some spatial symmetry.}

This bound on the energy is analogous to the well known (time symmetric) Penrose inequality \cite{Penrose:1973um,Bray:2003ns}, which states that the energy is always bounded from below by a function of the area of a minimal surface. The key difference is that in the Penrose inequality, the minimal surface has co-dimension one in the initial data. 

We start in Sec.~2 with {AdS${}_4$} initial data ($d=3$) with boundary $S^1\times S^1$.
We numerically explore a large class of solutions to the  constraint with a minimal $S^1$ of length $L$, and compute their energies. From this data, we determine a minimum energy, {$E_{min}(L)$}, for all solutions with a given $L$.
 There is one value of $L$ which corresponds to the AdS soliton and gives a  static solution. We compute $E_{min}(L)$ and find that it starts at zero, decreases to a minimum at the AdS soliton value, and then increases like $ L^3$ for large $L$.
 Even though the relative size of the two boundary circles, parameterized by $s$, is a conformal invariant labelling different boundary conditions, $E_{min}(L)$ depends on $s$ only through a simple rescaling. 
 {The} statement that $E\ge E_{min}(L)$ for initial data with a minimal circle of length $L$ can be viewed as a refined version of {the statement that the AdS soliton has minimum energy.} 

We next explore asymptotically {AdS${}_5$} solutions with spatial boundary $S^1 \times S^2$. There is again a one parameter family of such boundary conditions described by the relative size $s$ of the $S^1$ and $S^2$. Static solutions with these boundary conditions were found in \cite{Copsey:2006br}. It was shown that there is always a static solution where the $S^1$ is not contractible, and if $s$ is small enough, there are two solutions where it is, and there are minimal $S^2$'s. 
The solution with larger minimal sphere always has lower energy.
In Sec.~3 we consider time symmetric initial data with these boundary conditions. For any $s$, we investigate a large class of initial data with a minimal sphere of any size $A$.  We compute their energy and derive a curve $E_{min}(A)$ for several values of $s$. {$E_{min}(A)$} always grows for large $A$ like $s A^2$. For $s$ large enough that there are no static solutions with minimal spheres, $E_{min}(A)$ monotonically increases {for all $A$}. For smaller $s$, the shape of the curve $E_{min}(A)$ is determined by the areas of the two minimal surfaces, $A_1$ and $A_2 > A_1$, in the static solutions: $E_{min}(A)$
initially increases and has a local maximum at $A_1$, then decreases with a (possibly local) minimum at  $A_2$, and then increases again indefinitely. The minimum at $A_2$ is a global minimum except for a narrow range of $s$ of order one.

The fact that the extrema of the curve coincide with the static solutions (in both the AdS${}_4$ case and the AdS${}_5$ case) can be understood from the theorem that static solutions always extremize the energy \cite{Sudarsky:1992ty}. This means that we can obtain some information about the structure of $E_{min}(A)$ just by looking at the known static solutions.

Although our curves $E_{min}(L)$ and $E_{min}(A)$ are obtained numerically by sampling a large class of initial data, we believe they provide a good approximation to the true lower bound on the energy for a given size minimal surface.  If {a  metric}  contains more than one minimal surface, our bound applies to the one with smallest area -- we always refer to the globally minimal surface. Although we only consider the vacuum equations, it is plausible that these same bounds hold for gravity coupled to  matter satisfying the dominant energy condition.

We conclude in Sec.~4 with a discussion of  some open problems.

\section{Energy inequality in AdS\texorpdfstring{${}_4$}{4}}
Consider AdS${}_4$ with conformal boundary
\begin{equation}\label{eq:4Dmetricbdy}
    d s^2|_{\partial \mathcal{M}}=-d t^2+d\chi^2+d \theta^2,
\end{equation}
where $\theta$ has periodicity $2\pi$ and $\chi$ has periodicity $s$. Since this is a conformal boundary, only the relative size matters, and we have chosen a conformal frame where the $\theta$ circle has unit radius.

When the $\chi$ circle is smaller than the $\theta$ circle ($s<2\pi$), the ground state with this conformal boundary is given by the AdS soliton \cite{Horowitz:1998ha}, with the metric given by
\begin{equation}
    d s^2=\frac{1}{z^2}\left[-d t^2+\frac{d z^2}{f(z)}+f(z) d \chi^2+d\theta^2\right] ,
\end{equation}
where 
\begin{equation}
    f(z)=1-\frac{z^{3}}{z_0^{3}},\quad s=\frac{4\pi}{3} z_0,
\end{equation}
and we have set the AdS radius to one. In these coordinates, $z=0$ is the conformal boundary. Relative to the pure AdS solution, this AdS soliton has negative energy\footnote{We will drop factors of Newton's constant $G$ which should multiply all our expressions for the energy.}
\begin{equation}\label{eq:energy_soliton}
    E=-\frac{ s }{16 \pi z_0^3}\times 2\pi=-\frac{8\pi^3}{27 s^2},
\end{equation}
where the factor of $2\pi$ comes from the periodicity of the $\theta$ circle. For $s>2\pi$, the ground state is given by the same solution but with $\chi$ and $\theta$ swapped. In other words, the ground state for any $s$ is an AdS soliton where the smaller of the two circles pinches off in the bulk. 

The momentum constraint for time symmetric initial data is automatically satisfied, and the Hamiltonian constraint is given simply by
\begin{equation}\label{eq:init_cstr}
    \mathcal{R}=2\Lambda,
\end{equation}
where $\mathcal{R}$ is the Ricci scalar of the $d$-dimensional metric at $t=0$ and $\Lambda$ is 
the cosmological constant, in this case negative. 
One expects minimum energy solutions to have spatial symmetries. Since we want to minimize the  energy, we consider  time symmetric initial data with $U(1)^2$ symmetry. The most general such metric takes the form\footnote{We  also require that the size of the $\theta$ circle monotonically increases. This is also expected for minimum energy solutions.}
\begin{equation}\label{eq:4Dchoice}
    ds^2 = \frac{1}{z^2}\left[\frac{dz^2}{\a(z)\beta(z)}+ \a(z) d\chi^2  + d\theta^2\right],
\end{equation}
where $\a(z)$ has an asymptotic expansion
\begin{equation}\label{eq:4D_alpha_expand}
    \a (z) = 1 + O(z^3).
\end{equation}
There are infinitely many corner conditions that the asymptotic expansion of the initial data must satisfy if we want the full Lorentzian evolution to be $C^\infty$ \cite{Friedrich:1995vb,Enciso:2014lwa,Carranza:2018wkp,Horowitz:2019dym}. For example, the static boundary condition \eqref{eq:4Dmetricbdy} completely fixes the metric at orders $z$, $z^2$ and $z^4$, while the metric at order $z^3$ is free and determines the total energy. Imposing  \eqref{eq:4D_alpha_expand} only ensures that the energy is finite, but does not ensure the smoothness of the full Lorentzian metric. This means that our proposed energy inequality is not restricted to smooth spacetimes.

A nice feature of the choice of the ansatz \eqref{eq:4Dchoice} is that the Hamiltonian constraint takes the form
\begin{equation}\label{eq:beta}
\alpha\left(-6 \beta+2 z \beta^{\prime}\right)-\frac{1}{2} z\left(z \alpha^{\prime} \beta^{\prime}+\beta\left(-8 \alpha^{\prime}+2 z \alpha^{\prime\prime}\right)\right)=-6.
\end{equation}
Note that this is a first-order ODE for $\beta(z)$, so we can easily solve for $\beta(z)$  for any given $\a(z)$. 
Since there is only one asymptotic region, one of the circles must be contractible in the interior, and we are parameterizing that by $\chi$. So we want $\alpha(z)$ to vanish at some $z_0$.
The integration constant in the solution for $\beta(z)$ is fixed by requiring that the geometry be smooth at $z_0$ (no conical singularity), which is a relation between the periodicity $s$ and the metric components at $z_0$ and is given by 
\begin{equation}
    \beta(z_0)=\left(\frac{4 \pi}{s\,\a^{\prime}(z_0)}\right)^2.
\end{equation}
Note that at $z_0$, there is a minimal circle of length $L = 2\pi/z_0$.

The energy can be computed using the background subtraction method \cite{Hawking:1995fd}, where the background is chosen to be the Poincare patch of  AdS, compactified to a torus. With our ansatz \eqref{eq:4Dchoice}, it is given by
\begin{equation}
    E=  \frac{s}{8} (A_3 - 2 B_3),
\end{equation}
where $A_n$ and $B_n$ are the Taylor expansion coefficients of $\a$ and $\b$ in powers of $z$. The energy defined this way generally differs from the counter-term method \cite{Balasubramanian:1999re} by a constant, but this constant is zero in this case.

For numerical implementation, we take the following set of initial data:
\begin{equation}
    \a(z) = 1 + \sum_{n=3}^{n_0}  A_n z^n.
\end{equation}
Requiring $\a(z_0)=0$ puts a constraint on the $A_n$'s so that we have a {$(n_0 -3)$}-parameter family of functions. For a given $z_0>0$, we then use the following simple algorithm to compute $E_{min}$: 
\begin{enumerate}
    \item choose an initial $n_0$, a desired step size, an initial step size that is much larger, and an initial $\a(z)$ by choosing initial values of $A_n$ for $n=3,...,n_0-1$ such that $\a(z)$ is physical, i.e. $\a'(z_0)<0$  and $\a(z)>0$ for $z\in(0,z_0)$;
    \item generate a collection of functions $\alpha(z)$ by changing these $A_n$'s independently by a certain number or percentage (step size), throwing away those where $\a(z)$ becomes negative in the range $z\in(0,z_0)$;
    \item use the initial data constraint \eqref{eq:beta} to determine $\beta(z)$, throwing away those where $\beta(z)$ becomes zero or infinity anywhere in the range $z\in(0,z_0]$;
    \item for all remaining initial data, compute the energy for each and check if any of them has lower energy than the initial guess:
    \begin{itemize}
        \item if no: record the energy and $A_n$'s; 
        \item if yes: take the one with the lowest energy and feed back to step 1 with this new $\a(z)$;
    \end{itemize}
    \item take the recorded $A_n$'s and feed back to step 1 with a smaller step size if the desired step size is not reached, outputting the energy otherwise;
    \item repeat steps 1-5 until with a larger $n_0$ until the computed curve does not change significantly by increasing $n_0$.
\end{enumerate}

\begin{figure}
    \centering
    \includegraphics[width=0.55\linewidth]{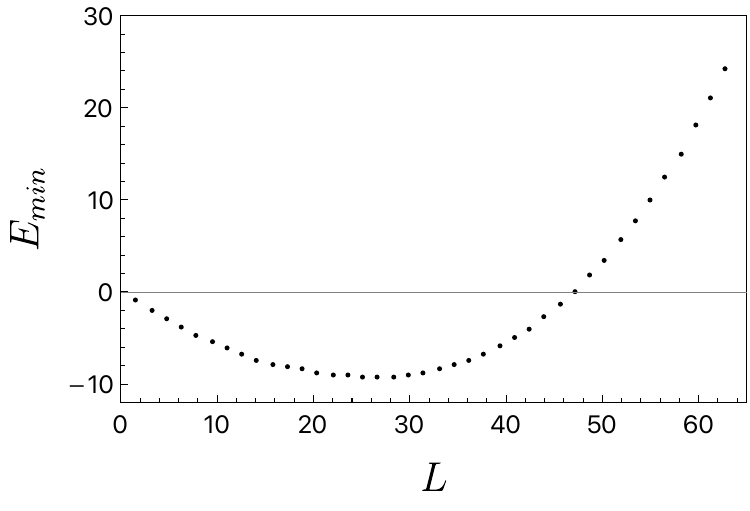}
    \caption{Minimum energy plot for asymptotically AdS${}_4$ solutions with toroidal conformal boundary, containing a minimal circle of length $L$. The length of the contractible circle is $s=1$, and the length of the noncontractible circle is $2\pi$. }
    \label{fig:AdS4s1}
\end{figure}

After doing this for different values of $z_0$, we obtain a plot of $E$ against $L=2\pi/z_0$. This is {our} approximation of $E_{min}(L)$. For $s=1$, the results are shown in Fig.~\ref{fig:AdS4s1}. We can see that the curve goes to zero as $L\rightarrow 0$ since there is no minimal circle in this limit and the Poincare patch of AdS (toroidally compactified) satisfies our conditions. $E_{min}(L)$ decreases from zero and reaches a minimum before increasing. The global minimum corresponds to the AdS soliton. 
At large $L$, we find that $E_{min}$ grows like $L^3$:
\be\label{eq:4Dasymp1}
E_{min} (L) \approx (2.4\times 10^{-4})\,  L^3 \qquad {\rm for\ large\ }L  .
\ee

An obvious guess for the function $E_{min}(L)$ is a cubic polynomial. Using $E_{min}'(L_*)=0$, $E_{min}(L_*)=E_*$, $E_{min}(0)=0$ and \eqref{eq:4Dasymp1}, where $L_*$ and $E_*$ are the values for the {AdS soliton}, we can determine this function completely and plot it. However, we find that there are data points lying below this naive guess. This means that the function cannot be a simple cubic polynomial.  

Having obtained the curve $E_{min}(L)$ for $s=1$, the corresponding functions for other values of $s$ can be obtained by the following scaling argument. Start with our minimal energy $s=1$ solution for some $L$, which has a conformally invariant ratio of the size of the $\chi$ to $\theta$ circles of $\sigma = 1/2\pi$. From this one solution, we can obtain solutions with any $\sigma$ by simply changing the periodicity of  $\theta$ to $2\pi/\lambda$. This multiplies $\sigma$ by $\lambda$, and changes $L$ by $1/ \lambda$. The energy gets two corrections: since gravitational energy is a surface integral at infinity, it is multiplied by $1/\lambda$. But to obtain the energy in the conformal frame of our standard boundary metric with $s=\lambda$, we have to rescale all boundary distances by a factor of $\lambda $ to restore the periodicity of $\theta$ to $2\pi$. Since energy is a dimensionful quantity, this multiplies the energy by another factor of $1/\lambda$. The net result is that, if we change the periodicity of $\theta$ we get:
\be\label{eq:scaling}
E_{s=\lambda}( L/\lambda) = E_{s=1} (L)/\lambda^2.
\ee
The solutions we obtain this way must be the minimum energy solutions with that $s$, since if there was a lower energy solution, one could reverse the argument and get a lower energy solution with $s=1$, contradicting our original bound.
In particular, since $E_{min}$ grows like $L^3$ for large $L$, \eqref{eq:scaling} implies that the scaling with $s$ is linear in this regime:
\be\label{eq:4Dasymp}
E_{min} (L) \approx (2.4\times 10^{-4})\, s L^3  \qquad {\rm for\ large\ }L .
\ee

\section{Energy inequality in AdS\texorpdfstring{${}_5$}{5}}
Now consider asymptotically AdS${}_5$ spacetimes with conformal boundary
\begin{equation}\label{eq:5Dmetricbdy}
    d s^2|_{\partial \mathcal{M}}=-d t^2+d\chi^2+d \Omega_2^2,
\end{equation}
where $d\Omega_2^2$ is the metric of a unit 2-sphere, and $\chi$ is periodically identified with period $s$. The spatial geometry is therefore a circle cross a sphere. We have chosen a conformal frame where the sphere is of unit radius, so $s$ parameterizes the size of the circle relative to the sphere. 

Static  solutions with this conformal metric have been studied numerically in \cite{Copsey:2006br}. There are two types of static solutions: either the sphere or the circle pinches off. There is a unique $S^2$-contractible static solution {regardless of what value $s$ takes}. In fact, this solution is independent of $s$: it exists when the $\chi$ direction is non-compact, and we can periodically identify $\chi$ everywhere with any periodicity. The $S^1$-contractible solutions, which have a minimal $S^2$, form a one-parameter family labelled by the area $A\in (0,\infty)$ of the minimal sphere. As $A$  increases, $s$ increases from 0 to some maximal value $s_{max}\approx 3.56$ before decreasing and asymptoting to zero in the large $A$ limit \cite{Copsey:2006br}; therefore, at any given $s<s_{max}$, there are two static $S^1$-contractible solutions, and none if $s>s_{max}$.

From the expectation that the ground state is given by a static solution with maximal symmetry allowed by the boundary conditions, for a given $s$, the {minimum energy solution} is either the static $S^2$-contractible solution or the lower-energy one of the two $S^1$-contractible solutions. The dividing point is $s=s_{crit}\approx 2.98$: below this value the ground state is a static solution with minimal $S^2$; above $s_{crit}$, it is the static $S^2$-contractible solution \cite{Copsey:2006br}. From the standpoint of gravitational holography, there is  a zero temperature quantum phase transition in the dual field theory as $s$ is varied across $s_{crit}$. 

We are interested in time symmetric initial data, which are generically non-static under time evolution. The most general time symmetric initial data with maximal symmetry allowed by the boundary conditions  \eqref{eq:5Dmetricbdy} takes the form
\begin{equation}\label{eq:AdS5_ansatz}
    ds^2 = \frac{1}{z^2}\left[ \frac{dz^2}{\a(z)\beta(z)}+ \a(z) d\chi^2 + d\Omega_2^2\right],
\end{equation}
where
\begin{equation}\label{eq:5Dasymp}
\begin{aligned}
    \a(z) &= 1 + \frac{z^2}{2}-\frac{1}{12}z^4\log z+A_4 z^4 + O(z^5,z^5\log z) ,\\
    \beta(z)&=1+\frac{z^2}{6}-\frac{1}{12}z^4\log z+B_4 z^4 +O(z^5,z^5\log z).
\end{aligned}
\end{equation}
As in the AdS${}_4$ case, we have imposed the minimum set of corner conditions to ensure the finiteness of the energy but no more. More specifically, we allow the asymptotic expansion of the function $\a(z)$ to be unconstrained starting from the order of the {energy}, $z^4$. 

Using \eqref{eq:AdS5_ansatz}, the Hamiltonian constraint \eqref{eq:init_cstr} is given by
\begin{equation}\label{eq:5DODE}
3 \alpha\left(-4 \beta+z \beta^{\prime}\right)-\frac{1}{2} z\left(z\left(-4+\alpha^{\prime} \beta^{\prime}\right)+2 \beta\left(-6 \alpha^{\prime}+z \alpha^{\prime\prime}\right)\right)=-12.
\end{equation}
As in the AdS${}_4$ case, this is a first-order ODE for the function $\beta(z)$. We note that the $\beta(z)$ expansion in \eqref{eq:5Dasymp} will be automatically ensured from solving \eqref{eq:5DODE} as long as the $\a(z)$ asymptotic expansion is imposed. We choose $\alpha(z)$ to vanish at $z_0$, so there is a minimal $S^2$ with $A = 4\pi/z_0^2$, and adjust the free constant in $\beta$ so there is no conical singularity.

Using the background subtraction method, the energy is given by
\begin{equation}\label{eq:E_def_5D}
    E=\frac{s}{4} (A_4 -3 B_4  +C_0),
\end{equation}
where $C_0$ is a constant that depends on the choice of the background. We choose the background to be the static $S^2$-contractible solution, in which case $C_0=-0.0348122$. Incidentally, the counter-term energy 
of \cite{Balasubramanian:1999re} is given by \eqref{eq:E_def_5D} with $C_0=-7/24$.

We then obtain an approximation for $E_{min}(A)$ using the algorithm outlined in the previous section by studying an $(n_0-4)$-parameter family of initial data:
\begin{equation}
    \a(z) = 1 + \frac{z^2}{2}-\frac{1}{12}z^4\log z+\sum_{n=4}^{n_0}  A_n z^n.
\end{equation}

\begin{figure}[H]
    \centering
    \begin{subfigure}[b]{.55\linewidth}
        \includegraphics[width=\textwidth]{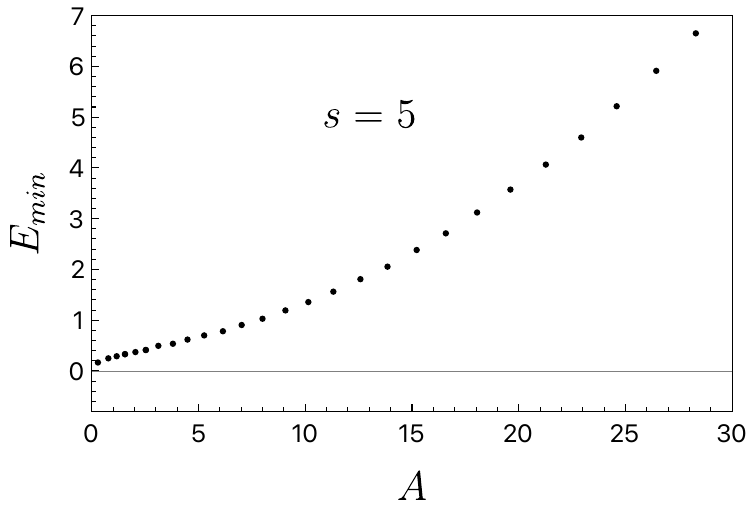} 
    \end{subfigure}
    \begin{subfigure}[b]{.55\linewidth}
        \includegraphics[width=\textwidth]{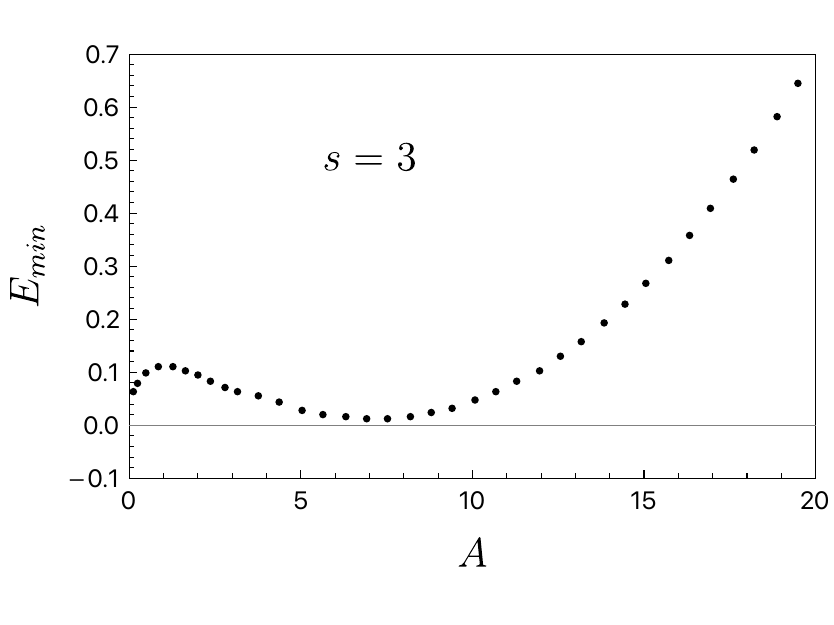}
    \end{subfigure}
    \begin{subfigure}[b]{.55\linewidth}
        \includegraphics[width=\textwidth]{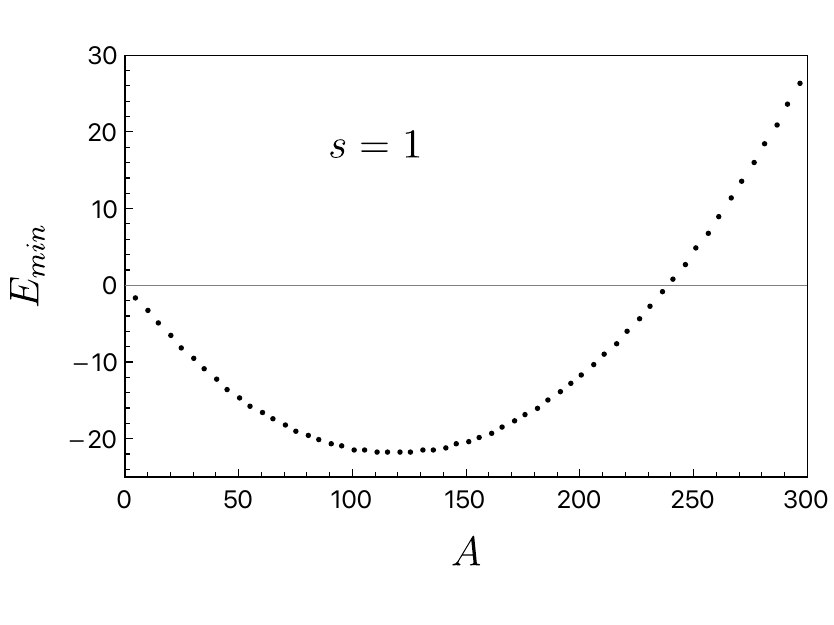}
    \end{subfigure}
    \caption{Minimum energy plots for asymptotically AdS${}_5$ solutions with boundary $S^1$ (with length $s$) times a unit  $S^2$, containing a minimal $S^2$ with area $A$.}
    \label{fig:S1S2s135}
\end{figure}

Fig.~\ref{fig:S1S2s135} displays the resulting curves for $s=5,3,1$. Since static solutions extremize the energy, we expect extrema of $E_{min}(A)$ at those $A$'s where there is a static solution with a minimal sphere with that area. For $s=5$, which is greater than $s_{max}$, there is no static solution with minimal $S^2$ and the curve monotonically increases. For $s=3$, which is just above $s_{crit}\approx 2.98$ but below $s_{max}$, $E_{min}(A)$  initially increases and has a local maximum at the static solution with the smaller minimal sphere,  then drops down to the small positive energy of the static solution with the larger minimal sphere, before increasing again.  As we decrease $s$, the energy of both extrema decrease, with the first approaching zero and the second becoming negative.
For $s=1$, the initial local maximum is {at $A \approx0.06$, which is} below our resolution, so we only see the curve decrease to a global minimum at the static solution with larger minimal $S^2$, before increasing.

{It is difficult to obtain reliable values of $E_{min}(A)$ close to $A=0$ since the curvature grows as $A \to 0$.} However, we expect the {curve} to go to zero in the limit $A$ goes to zero. This is because in this limit there is no minimal $S^2$, so the ground state should be the static solution with $S^2$ contractible. This is precisely our reference background.

\begin{figure}[H]
    \centering
    \includegraphics[width=0.5\linewidth]{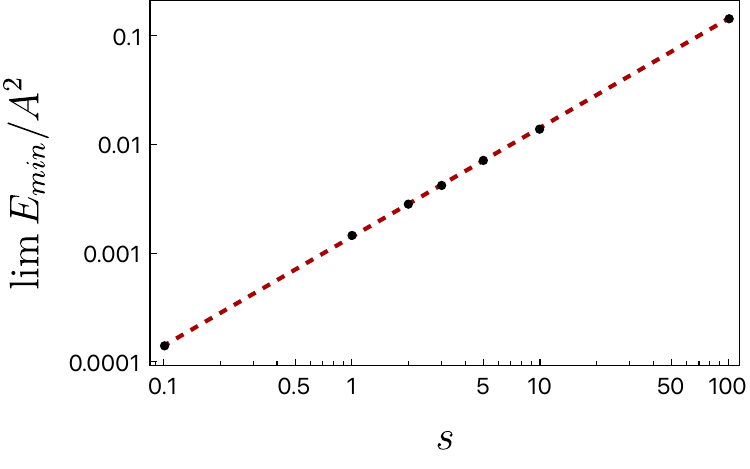}
    \caption{Dependence {on $s$} of the quadratic coefficient of $E_{min}(A)$ in the large $A$ limit,  for asymptotically AdS${}_5$ solutions.}
    \label{fig:AdS5_quad}
\end{figure}

At large $A$, we find that the energy scales like $A^2$ for any value of $s$. The $s$ dependence can be found by computing the large $A$ limit of $E_{min}/A^2$, as shown in Fig.~\ref{fig:AdS5_quad}. We find that the large $A$ behavior of $E_{min}$ is given by
\begin{equation}\label{eq:linear2}
    E_{min}(A) \approx (1.4\times 10^{-3})\, s A^2.
\end{equation}

The fact that the large $A$ behavior of $E_{min}(A)$ is linear in $s$ can be understood by the following analog of the scaling argument at the end of Sec.~2. For large $A$, the curvature of the $S^2$ is negligible, and one should obtain the same function $E_{min}(A)$ for $S^1\times T^2$ boundary conditions with the contractible $S^1$ having period $s$ and the $T^2$ having area $4\pi$. Starting with a minimum energy solution with $s=1$, and minimal $T^2$ with area $A$, change the periodicity of $T^2$ so that the boundary torus has area $4\pi/\lambda^2$. This changes the area of the minimal $T^2$ by $1/\lambda^2$. The energy again gets two corrections. It gets a factor of $1/\lambda^2$ since it is a surface integral, and it gets another factor of $1/\lambda$ since we have to rescale all boundary distances by $\lambda$ to put the boundary metric in our standard conformal frame where the area of $T^2$ is $4\pi$. This constant rescaling changes $s=1$ to $s=\lambda$. So the net result is that, if we change the periodicity of $T^2$, we get 
\be\label{eq:scaling2}
E_{s=\lambda}( A/\lambda^2) = E_{s=1} (A)/\lambda^3  \qquad {\rm for \ }S^1\times T^2 \rm{\ boundary} .
\ee
Since $E_{min}(A)$ {should satisfy this for large $A$ where it} is proportional to $A^2$, {\eqref{eq:scaling2} implies that} $E_{min}(A)$ is linear in $s$ as in \eqref{eq:linear2}.

\section{Discussion}

We have numerically computed a bound on the energy of time symmetric, asymptotically AdS${}_4$ or AdS${}_5$ initial data in terms of the area of a co-dimension two minimal circle or sphere. 

This work can be extended in several directions. First there is the obvious extension to higher dimensional AdS with spatial boundary $S^1\times S^n$, but we expect the results to be similar. More interestingly, one could consider boundaries $S^1\times \Sigma$ where $\Sigma$ is a general Riemannian manifold, and bound the energy in terms of minimal surfaces that are not topological spheres. Alternatively, one could consider the case when the boundary is a product of spheres (no spatial circle). Then one could get bounds on the energy in terms of higher codimension minimal surfaces.

Perhaps the most important open question is to derive an analytic form of our bounds $E_{min}(L)$ and $E_{min}(A)$ (or any of the above generalizations) and prove that they provide lower bounds on the energy. 
In the AdS${}_4$ case, we established that $E_{min}(L)$ is not a simple cubic polynomial. This is not surprising as the Penrose inequality generally involves non-integer powers \cite{Itkin:2011ph}. It is likely that non-integer powers are present in this new inequality, too. Since we have (numerically) found the coefficient of the cubic term, it is then feasible to subtract the cubic term and make a log-log plot for large $L$ to determine the next-to-leading order power. If the actual analytic expression is not too complicated, e.g., if it is just a sum of a small number of fractional powers, this procedure may be enough to extract the form of the function. A similar procedure can be done for the AdS${}_5$ case: the curve is more complicated, but we also have more information given that the static solutions are local extrema. 

Another open question is whether there is a unique {solution}  at every point on the curves $E_{min}(L)$ and $E_{min}(A)$. The usual positive energy theorem not only says $E\ge 0$ but also specifies that the ground state is unique. In our case, we do not have evidence for  uniqueness. In fact, even though we have initial data with energy very close to the lower bound, we have not  shown the existence of solutions that saturate the bound, except at the extrema where they are given by the static solutions.

The new energy inequality is a refinement of the positive energy theorem in AdS. In the positive energy theorem, only the true ground state matters, but here all static solutions are important as they tell us about turning points of the curve. When the spatial boundary is a product of spheres, the number of static solutions  with minimal spheres can be arbitrarily large depending on the relative size of the spheres \cite{Horowitz:2022hlz}. We then expect the curve to have a correspondingly large number of turning points.

It is natural to ask whether there are analogous bounds in the asymptotically flat case. Since we require that the asymptotic boundary contain a circle, the appropriate context is (five-dimensional) Kaluza-Klein theory. It is known that there is no positive energy theorem for this theory \cite{Witten:1981gj}.\footnote{One can restore positivity of energy by adding fermions and requiring periodic boundary conditions around the asymptotic circle.} There are solutions with minimal $S^2$'s and arbitrarily negative energy \cite{Brill:1989di,Brill:1991qe}. These solutions are often called ``bubbles of nothing" since, after dimensional reduction on the $S^1$, space resembles Euclidean space with a ball removed. The minimal $S^2$ is the boundary of this ball, and there is nothing inside. It was shown in \cite{Brill:1989di} that the energy must always be greater than a certain ``radius" of the bubble, which is not directly related to the proper area $A$ of the minimal $S^2$. It would be interesting to check whether there is a lower bound in terms of $A$.

The existence of our bound $E_{min}$ is likely to have physical implications for holography. One expects every geometric property of the bulk to correspond to some property of the dual field theory (in the large $N$ limit). Our result would then bound the energy 
of holographic field theories with this property. In particular, there must be new states corresponding to wrapping branes around the minimal surface.

\section*{Acknowledgements} 
It is a pleasure to thank Jingbo Wan for discussions. GH  is supported in part by NSF grant PHY-2107939. DW is partly supported by NSF grants PHY-2107939 and PHY-2207659. DW is grateful for the hospitality of the Aspen Center for Physics, where part of this work was performed, funded by NSF grant PHY-2210452.

\bibliographystyle{JHEP}
\bibliography{library}

\providecommand{\href}[2]{#2}\begingroup\raggedright\begin{thebibliography}{10}

\bibitem{Schoen:1979zz}
R.~Schoen and S.-T. Yau, {\it {Positivity of the Total Mass of a General
  Space-Time}},  {\em Phys. Rev. Lett.} {\bf 43} (1979) 1457--1459.

\bibitem{Witten:1981mf}
E.~Witten, {\it {A New Proof of the Positive Energy Theorem}},  {\em Commun.
  Math. Phys.} {\bf 80} (1981) 381.

\bibitem{Gibbons:1983aq}
G.~W. Gibbons, C.~M. Hull, and N.~P. Warner, {\it {The Stability of Gauged
  Supergravity}},  {\em Nucl. Phys. B} {\bf 218} (1983) 173.

\bibitem{Wang:2001}
X.~Wang, {\it {Mass for asymptotically hyperbolic manifolds}},  {\em J. Diff.
  Geom.} {\bf 57} (2001) 273--299.

\bibitem{Maerten:2006}
D.~Maerten, {\it {Positive energy-momentum theorem for AdS-asymptotically
  hyperbolic manifolds}},  {\em Ann. Henri Poincare} {\bf 7} (2006) 975--1011.

\bibitem{Horowitz:1998ha}
G.~T. Horowitz and R.~C. Myers, {\it {The AdS / CFT correspondence and a new
  positive energy conjecture for general relativity}},  {\em Phys. Rev. D} {\bf
  59} (1998) 026005, [\href{http://arxiv.org/abs/hep-th/9808079}{{\tt
  hep-th/9808079}}].

\bibitem{Brendle:2024}
S.~Brendle and P.-K. Hung, {\it {Systolic inequalities and the Horowitz-Myers
  conjecture}},  \href{http://arxiv.org/abs/2406.04283}{{\tt
  arXiv:2406.04283}}.

\bibitem{Penrose:1973um}
R.~Penrose, {\it {Naked singularities}},  {\em Annals N. Y. Acad. Sci.} {\bf
  224} (1973) 125--134.

\bibitem{Bray:2003ns}
H.~L. Bray and P.~T. Chrusciel, {\it {The Penrose inequality}},
  \href{http://arxiv.org/abs/gr-qc/0312047}{{\tt gr-qc/0312047}}.

\bibitem{Copsey:2006br}
K.~Copsey and G.~T. Horowitz, {\it {Gravity dual of gauge theory on S**2 x S**1
  x R}},  {\em JHEP} {\bf 06} (2006) 021,
  [\href{http://arxiv.org/abs/hep-th/0602003}{{\tt hep-th/0602003}}].

\bibitem{Sudarsky:1992ty}
D.~Sudarsky and R.~M. Wald, {\it {Extrema of mass, stationarity, and staticity,
  and solutions to the Einstein Yang-Mills equations}},  {\em Phys. Rev. D}
  {\bf 46} (1992) 1453--1474.

\bibitem{Friedrich:1995vb}
H.~Friedrich, {\it {Einstein equations and conformal structure - Existence of
  anti de Sitter type space-times}},  {\em J. Geom. Phys.} {\bf 17} (1995)
  125--184.

\bibitem{Enciso:2014lwa}
A.~Enciso and N.~Kamran, {\it {Lorentzian Einstein metrics with prescribed
  conformal infinity}},  {\em J. Diff. Geom.} {\bf 112} (2019), no.~3 505--554,
  [\href{http://arxiv.org/abs/1412.4376}{{\tt arXiv:1412.4376}}].

\bibitem{Carranza:2018wkp}
D.~A. Carranza and J.~A. Valiente~Kroon, {\it {Construction of anti-de
  Sitter-like spacetimes using the metric conformal Einstein field equations:
  the vacuum case}},  {\em Class. Quant. Grav.} {\bf 35} (2018), no.~24 245006,
  [\href{http://arxiv.org/abs/1807.04212}{{\tt arXiv:1807.04212}}].

\bibitem{Horowitz:2019dym}
G.~T. Horowitz and D.~Wang, {\it {Gravitational Corner Conditions in
  Holography}},  {\em JHEP} {\bf 01} (2020) 155,
  [\href{http://arxiv.org/abs/1909.11703}{{\tt arXiv:1909.11703}}].

\bibitem{Hawking:1995fd}
S.~W. Hawking and G.~T. Horowitz, {\it {The Gravitational Hamiltonian, action,
  entropy and surface terms}},  {\em Class. Quant. Grav.} {\bf 13} (1996)
  1487--1498, [\href{http://arxiv.org/abs/gr-qc/9501014}{{\tt gr-qc/9501014}}].

\bibitem{Balasubramanian:1999re}
V.~Balasubramanian and P.~Kraus, {\it {A Stress tensor for Anti-de Sitter
  gravity}},  {\em Commun. Math. Phys.} {\bf 208} (1999) 413--428,
  [\href{http://arxiv.org/abs/hep-th/9902121}{{\tt hep-th/9902121}}].

\bibitem{Itkin:2011ph}
I.~Itkin and Y.~Oz, {\it {Penrose Inequality for Asymptotically AdS Spaces}},
  {\em Phys. Lett. B} {\bf 708} (2012) 307--308,
  [\href{http://arxiv.org/abs/1106.2683}{{\tt arXiv:1106.2683}}].

\bibitem{Horowitz:2022hlz}
G.~T. Horowitz, D.~Wang, and X.~Ye, {\it {An infinity of black holes}},  {\em
  Class. Quant. Grav.} {\bf 39} (2022), no.~22 225014,
  [\href{http://arxiv.org/abs/2206.08944}{{\tt arXiv:2206.08944}}].

\bibitem{Witten:1981gj}
E.~Witten, {\it {Instability of the Kaluza-Klein Vacuum}},  {\em Nucl. Phys. B}
  {\bf 195} (1982) 481--492.

\bibitem{Brill:1989di}
D.~Brill and H.~Pfister, {\it {States of Negative Total Energy in
  {Kaluza-Klein} Theory}},  {\em Phys. Lett. B} {\bf 228} (1989) 359--362.

\bibitem{Brill:1991qe}
D.~Brill and G.~T. Horowitz, {\it {Negative energy in string theory}},  {\em
  Phys. Lett. B} {\bf 262} (1991) 437--443.

\end{thebibliography}\endgroup

\end{document}